\documentclass[12pt,reqno]{amsart}

\usepackage{amssymb}
\usepackage{amsaddr}

\usepackage{lineno}

\usepackage{amsmath,amsthm,amssymb,amsfonts}
\usepackage{bm} 
\usepackage{upgreek} 
\usepackage{pifont} 
\usepackage{mathdots}
\usepackage{subcaption}
\usepackage{graphicx}

\usepackage{natbib}
\setcitestyle{square,numbers,sort&compress}

\usepackage[colorlinks,linkcolor=red,anchorcolor=blue,citecolor=green]{hyperref}
\usepackage[capitalise,noabbrev,sort&compress]{cleveref}

\newtheorem{theorem}{Theorem}

\newtheorem{corollary}[theorem]{Corollary}
\newtheorem{definition}[theorem]{Definition}
\newtheorem{example}[theorem]{Example}
\newtheorem{lemma}[theorem]{Lemma}

\newtheorem{problem}[theorem]{Problem}
\newtheorem{remark}[theorem]{Remark}

\crefname{equation}{}{}

\newcommand{\CC}{\mathbb{C}}

\newcommand{\NN}{\mathbb{N}}

\newcommand{\RR}{\mathbb{R}}

\newcommand{\cA}{\mathcal{A}}

\newcommand{\cI}{\mathcal{I}}

\newcommand{\cK}{\mathcal{K}}

\newcommand{\cM}{\mathcal{M}}

\newcommand{\cT}{\mathcal{T}}

\newcommand{\fA}{\mathfrak{A}}

\newcommand{\fX}{\mathfrak{X}}
\newcommand{\fY}{\mathfrak{Y}}

\newcommand{\Lalgsl}{\mathfrak{sl}}
\newcommand{\Lalggl}{\mathfrak{gl}}

\newcommand{\ii}{{\mathrm{i}}}
\newcommand{\textif}{\text{if }}
\newcommand{\otherwise}{\text{otherwise}}

\newcommand{\paren}[1]{\left( #1 \right)}
\newcommand{\brac}[1]{\left[ #1 \right]}
\newcommand{\set}[1]{\left\{ #1 \right\}}
\newcommand{\lrangle}[1]{\left\langle #1 \right\rangle}

\newcommand{\cond}[1]{\,\left( #1 \right)}

\newcommand{\card}[1]{\left\lvert #1 \right\rvert}
\newcommand{\abs}[1]{\left\lvert #1 \right\rvert}

\newcommand{\lrang}[1]{\left\langle #1 \right\rangle}

\newcommand{\bra}[1]{\left\langle #1 \right|}
\newcommand{\ket}[1]{\left| #1 \right\rangle}

\newcommand{\wt}[1]{\widetilde{#1}}
\newcommand{\ol}[1]{\overline{#1}}

\DeclareMathOperator{\Tr}{Tr}

\DeclareMathOperator{\diag}{diag}

\DeclareMathOperator{\Span}{span}

\DeclareMathOperator{\Aut}{Aut}
\DeclareMathOperator{\Sym}{Sym}
\DeclareMathOperator{\OW}{OW}
\DeclareMathOperator{\Ad}{Ad}

\date{}

\begin{document}

\title{Quantum walks on simplexes and multiple perfect state transfer}

\author{Hiroshi Miki}
\address{Department of Mathematical Sciences, Faculty of Science and Engineering, Doshisha University, Kyotanabe City, Kyoto, Japan}
\email{hmiki@mail.doshisha.ac.jp}

\author{Satoshi Tsujimoto}
\address{Graduate School of Informatics,
	Kyoto University, Sakyo-Ku, Kyoto, 606 8501, Japan}
\email{tsujimoto.satoshi.5s@kyoto-u.jp}

\author{Da Zhao}
\address{Graduate School of Informatics,
	Kyoto University, Sakyo-Ku, Kyoto, 606 8501, Japan}
\email{zhao.da.77r@st.kyoto-u.ac.jp}

\begin{abstract}
	In this paper, we study quantum walks on the extension of association schemes.
	Various state transfers can be achieved on these graphs, such as multiple state transfer among extreme points of a simplex, fractional revival on subsimplexes.
	Since only few examples of multiple (perfect) state transfer are known, we aim to make some additions in this collection.
\end{abstract}


\maketitle

\section{Introduction}

Given a graph $\Gamma = (X, E)$, which might be weighted or oriented, the continuous-time quantum walk on this graph is given by
$U(t) = \exp(-\ii t A)$, where $A$ is a Hermitian matrix associated to the graph $\Gamma$.
This notion was raised by Farhi and Guttman \cite{farhi_quantum_1998} to develop quantum algorithms.
Bose \cite{bose_quantum_2003} investigated quantum walk on a path for the study of quantum information transmission.

Let $\ket{\bm{e}_x}$ be the characteristic vector for the vertex $x \in X$, and let $V = \oplus_{x \in X} \CC \ket{\bm{e}_x}$ be the space on which $U(t)$ acts.
Suppose one starts with a state, say $\ket{\bm{e}_u}$, after evolution by time $t$, one reaches
\begin{equation}
	U(t) \ket{\bm{e}_u} = \sum_{x \in S} c_x \ket{\bm{e}_x},
\end{equation}
where $S$ is a subset of $X$ and $\sum_{x \in S} \abs{c_x}^2 = 1$.
We say fractional revival (FR) occurs on $S$ at time $t$.
Several special cases have been focused in the research.
\begin{itemize}
	\item Generation of maximal entanglement (GME). \\
	      If $S = \set{v,w}$ and $\abs{c_v} = \abs{c_w} = \frac{1}{\sqrt{2}}$ for $x \in S$, then the resulting state at time $t$ is a maximally entangled state of $\bm{e}_v$ and $\bm{e}_w$.
	\item Perfect state transfer (PST). \\
	      If $S = \set{v}$, in other words there exists a real constant $\gamma$ such that $U(t) \ket{\bm{e}_u} = e^{\ii \gamma} \ket{\bm{e}_v}$, then we say that there exists perfect state transfer from $u$ to $v$ at time $t$.
	\item Pretty good state transfer (PGST). \\
	      If there exists a real sequence $\set{t_k}$ and a constant $\gamma \in \RR$ such that $\lim_{k \to \infty} U(t_k) \ket{\bm{e}_u} = e^{\ii \gamma} \ket{\bm{e}_v}$, then we say that there exists pretty good state transfer from $u$ to $v$.
	\item Multiple (perfect/pretty good) state transfer (MST) and universal state transfer (UST). \\
	      Let $C$ be a subset of $X$.
	      If for every $u, v \in C$, there exists perfect/pretty good state transfer from $u$ to $v$, then we say that there exists multiple perfect state transfer/multiple pretty good state transfer in $C$.
	      In particular, if $C = X$, then we say that there exists universal state transfer in the graph $\Gamma$.
	\item Zero transfer (ZT). \\
	      If there exists $v \in X$ such that for every $t \in \RR$, $\bra{\bm{e}_v} U(t) \ket{\bm{e}_u} = 0$, then we say that there is zero transfer between $u$ and $v$.
	      In other words, the state on $u$ can never be seen on $v$ at any time.
\end{itemize}

After Bose, the study turns to general graphs as perfect state transfer between antipodal points of an unweighted path only occurs when the length is $2$ or $3$.
Christandl et al.~\cite{christandl_perfect_2005} showed that perfect state transfer can be achieved between antipodal points of a hypercube of arbitrary dimension, which gives the perfect state transfer between the two ends of a weighted path through projection.
As perfect state transfers are hard to obtain, the notation of pretty good state transfer was introduced by Godsil \cite{godsil_state_2012}.
It was shown by Kay \cite{kay_basics_2011} that in an unoriented graph, if one fixes the initial state on $u$, then one can only achieve perfect state transfer between two vertices $u$ and $v$, but not a third vertex.
Recently Chaves et al. \cite{chaves_why_2022} discussed `Why and how to add direction to a quantum walk'.
They also studied the phenomenon of zero transfer.
The ambitious universal state transfer was constructed by Cameron et al. \cite{cameron_universal_2014} in oriented graphs, which is impossible for unoriented graphs.
Graphs that carry universal perfect state transfer and universal pretty good state transfer are partially constructed and characterized in \cite{cameron_universal_2014,connelly_universality_2017}.
However, these graphs are dense in the sense that almost every pair of vertices are adjacent.
Multiple state transfer, as a relaxation of universal state transfer, was proposed by Godsil and Lato in \cite{godsil_perfect_2020}.

In this paper, we focus on multiple perfect state transfer (MPST).
Since very few examples are known and the distance of transfer in known examples is small, we aim to add new models that will give multiple perfect state transfer with arbitrary length.
In order to achieve that end, we employ the techniques of commutative association schemes. 
More specifically, from the small graph where MPST can be observed, we will show that we can obtain the model that provide MPST by considering the symmetric tensor product of the underlying association schemes. 
Furthermore, the multivariate Krawtchouk polynomials play an important in explicit calculations.  
For example, in~\cite{godsil_state_2012} Godsil and Lato showed that the directed 3-gons carry multiple perfect state transfer. 
Since the directed 3-gon is an adjacency graph of a commutative association scheme, we obtain multiple perfect state transfer on triangle simplex of arbitrary length by taking symmetric tensor product of the association scheme.

The paper is organized as follows.

In \cref{sec:aerator}, we introduce the definition and basic properties of association schemes.
The extension of association schemes is emphasized.
In \cref{sec:overdose}, we analyze quantum walks on extension of association schemes.
In \cref{sec:incompliant}, we apply the analysis in \cref{sec:overdose} to several base association schemes, which gives various examples of state transfer, such as perfect state transfer among extreme points of a simplex, multiple perfect state transfer among extreme points of a simplex, and fractional revival on subsimplexes of a simplex.

\section{Association scheme}\label{sec:aerator}

In this section, we recall the definition and basic properties of commutative association scheme.
We will emphasize the extension of association scheme, which will be used in later construction of state transfers.

\subsection{Commutative association scheme}\label{sec:centenarianism}

A commutative association scheme $\mathfrak{X} = (X, \set{R_i}_{i=0}^d)$ on $X$ of class $d$ consists of the relations $R_i \cond{0 \leq i \leq d}$ whose adjacency matrices $A_i \cond{0 \leq i \leq d}$ satisfy the following conditions.
\begin{enumerate}
	\item $A_0 = I$;
	\item $A_0 + A_1 + \cdots + A_d = J$;
	\item $A_i^\top = A_{i'}$ for some $i' \in \set{0,1,\ldots, d}$;
	\item $A_i A_j = A_j A_i = \sum_{k=0}^d p_{i,j}^k A_k$.
\end{enumerate}

The Bose-Mesner algebra $\fA = \lrang{A_0, A_1, \ldots, A_d}$ of a commutative association scheme $\mathfrak{X} = (X, \set{R_i}_{i=0}^d)$ has another basis $\fA = \lrang{E_0, E_1, \ldots, E_d}$ composed of primitive idempotents.
They satisfy the following properties.
\begin{enumerate}
	\item $E_0 = \frac{1}{\card{X}} J$;
	\item $E_0 + E_1 + \cdots + E_d = I$;
	\item $E_i^\top = \overline{E_i} = E_{\hat{i}}$ for some $\hat{i} \in \set{0,1,\ldots, d}$;
	\item $E_i \circ E_j = \frac{1}{\card{X}} \sum_{k=0}^d q_{i,j}^k E_k$
\end{enumerate}

\begin{example}\label{empl:upsnatch}
	Let $A_0 =
		\begin{bmatrix}
			1 &   \\
			  & 1
		\end{bmatrix}$ and
	$A_1 =
		\begin{bmatrix}
			  & 1 \\
			1 &
		\end{bmatrix}$.
	The primitive idempotents are given by
	$E_0 = \frac{1}{2}\begin{bmatrix}
			1 & 1 \\
			1 & 1
		\end{bmatrix}$ and
	$E_1 = \frac{1}{2}\begin{bmatrix}
			\phantom{+}1 & -1           \\
			-1           & \phantom{+}1
		\end{bmatrix}$.
	Then $A_0 E_0 = E_0$, $A_0 E_1 = E_1$, $A_1 E_0 = E_0$, and $A_1 E_1 = - E_1$.
	In other words $A_j E_{\lambda} = (-1)^{j \cdot \lambda} E_{\lambda}$ for $j, \lambda \in \set{0,1}$.
	This association scheme is called the trivial association scheme $\mathfrak{X}_2$ of size 2.
	Indeed this association scheme is the underlying association scheme of a path of length $1$.
\end{example}

\begin{example}\label{empl:hoove}
	Let $Z = \begin{bmatrix}
			  & 1 &   &        &   \\
			  &   & 1 &        &   \\
			  &   &   & \ddots &   \\
			  &   &   &        & 1 \\
			1 &   &   &        &
		\end{bmatrix}$ be the circulant matrix of size $n$.
	Let $A_k = Z^k$ for $k = 0,1, \ldots, n-1$ and let $E_{\ell} = (\frac{\zeta^{\ell(j-k)}}{\sqrt{n}})_{0 \leq j,k \leq n-1}$, where $\zeta$ is the $n$-th root of unity.
	Then $A_k E_{\ell} = \zeta^{k \ell} E_{\ell}$ for $k, \ell \in \set{0,1,\ldots, n-1}$.
	This association scheme is called the association scheme $\mathfrak{Z}_n$ of directed $n$-gon.
\end{example}

Let $\mathfrak{X} = (X, \set{R_i}_{i=0}^d)$ be an association scheme on $X$ of class $d$.
Let $A_i \cond{0 \leq i \leq d}$ be the adjacency matrices of $\mathfrak{X}$, and let $E_i \cond{0 \leq i \leq d}$ be the primitive idempotents of $\mathfrak{X}$.
The transition matrices, the first eigenmatrix $P = (P_{i,j})_{0 \leq i, j \leq d}$ and the second eigenmatrix $Q = (Q_{i,j})_{0 \leq i, j \leq d}$, between the two bases are given by
\begin{equation}
	A_i = \sum_{j=0}^d P_{j,i} E_j, \qquad E_i = \frac{1}{\abs{X}} \sum_{j=0}^d Q_{j,i} A_j
\end{equation}
In particular we have the valencies $k_i = P_{0,i}$ and the multiplicities $m_i = Q_{0,i}$.
The two eigenmatrix are related by $PQ = \abs{X} I$ and $\frac{P_{i,j}}{k_j} = \frac{\overline{Q_{j,i}}}{m_i}$.
We denote the cosine matrix of these values by $C = (c_{i,j})_{0 \leq i,j \leq d} = \paren{\frac{P_{i,j}}{k_j}}_{0 \leq i,j \leq d}$.
The transition between these two bases is very useful in our computation.
The reader is referred to~\cite{bannai_algebraic_1984} for a thorough discussion of association scheme.

\subsection{Extension of association schemes}

In this subsection, we exhibit two operations to construct new association schemes from old ones.
The first operation is based on tensor product and the second operation is based on group action.
Combining these two operations, we obtain the so-called extension of association schemes, also known as the symmetric tensor product of association schemes~\cite{MR0384310}.

Suppose $\mathfrak{X} = (X, \set{R_i}_{i=0}^d)$ is an association scheme on $X$ of class $d$ and $\mathfrak{Y} = (Y, \set{S_i}_{i=0}^e)$ is an association scheme on $Y$ of class $e$.
Let $A_i \cond{0 \leq i \leq d}$ and $B_j \cond{0 \leq j \leq e}$ be their adjacency matrices respectively.
Then
\[
	A_i \otimes B_j \cond{0 \leq i \leq d, 0 \leq j \leq e}
\]
gives an association scheme on $X \times Y$ of class $(d+1)(e+1) -1$, denoted by $\fX \otimes \fY$.
In particular one can take tensor product of an association scheme $\fX$ with itself, which gives $\fX^{\otimes 2} = \fX \otimes \fX$.
By composition and associativity, we can define $\fX^{\otimes N} = \fX^{\otimes (N-1)} \otimes \fX = \fX \otimes \cdots \otimes \fX$.
The Bose-Mesner algebra of $\fX^{\otimes N}$ is denoted by $\cA^{\otimes N}$.

Let $G \leq \Aut(\mathcal{A})$ be a subgroup of the automorphisms of the Bose-Mesner algebra $\cA$ of a commutative association scheme $\fX$.
Then the $G$-invariant matrices in $\mathcal{A}$ form a Bose-Mesner algebra of a fusion scheme (also called subscheme) of $\fX$.

Let us consider the group action of the symmetric group $S_N$ on the Bose-Mesner algebra of $\fX^{\otimes N}$ given by
\begin{equation*}
	g(A_{i_1} \otimes \cdots \otimes A_{i_N}) = A_{i_{g(1)}} \otimes \cdots \otimes A_{i_{g(N)}}
\end{equation*}
for every $g \in S_N$. 
Here $A_{i_1}, A_{i_2}, \ldots, A_{i_N}$ are taken from $A_0, A_1, \ldots, A_d$. 
It can be easily verified that these actions give automorphisms of $\cA^{\otimes N}$.
The fusion scheme, denoted by $\Sym(\fX,N)$, is called the extension of $\fX$, also known as the $N$-th symmetric tensor product of $\fX$. 
A typical relation in $\Sym(\fX,N)$ is given by 
\begin{equation}
	\frac{1}{\binom{N}{m_0 ,m_1, \ldots ,m_d}} \sum_{\sigma \in S_N} A_{i_{\sigma(1)}} \otimes A_{i_{\sigma(2)}} \otimes \cdots \otimes A_{i_{\sigma(N)}}, 
\end{equation}
where $m_i$ is the number of occurrences of $A_i$ in $A_{i_1}, A_{i_2}, \ldots, A_{i_N}$ for $i = 0,1, \ldots, d$. 
Here $\binom{N}{m_0 ,m_1, \ldots ,m_d}$ is the multinomial number define by 
\begin{equation}
\binom{N}{m_0 ,m_1, \ldots ,m_d} = \frac{N!}{m_0! m_1! \cdots m_d!}.
\end{equation}

\begin{example}
	Recall that $\mathfrak{X}_2$ is the trivial association scheme of size $2$. 
	The binary Hamming association scheme $H(N,2)$ is the $N$-th symmetric tensor product of $\mathfrak{X}_2$.
\end{example}

Both tensor product and symmetric tensor product can be used to construct new association schemes with multivariate polynomial structure, see~\cite{bannai2023multivariate,bernard2022bivariate} for latest development on bivariate polynomial association schemes.

Let $\cI_{N, d} = \set{n = (n_0, n_1, \ldots, n_d) \in \NN_0^{d+1} \mid n_0 + n_1 + \cdots + n_d = N}$ be the set of multi-indices of length $d+1$ with sum $N$.
We adopt the multi-index notation in this paper.
For example, if $p = (p_0, p_1, \ldots, p_d)$ and $n = (n_0, n_1, \ldots, n_d)$, then $\wt{p}^{n} = \wt{p}_0^{n_0} \cdots \wt{p}_d^{n_d}$, $n! = n_0! n_1! \cdots n_d!$, and $\binom{N}{n} = \frac{N!}{n!}$ whence $N = n_0 + n_1 + \cdots + n_d$.

Consider the $N$-th symmetric tensor product $\Sym(\fX,N)$ of $\mathfrak{X}$.
We denote by $\alpha = (\alpha_0, \alpha_1, \ldots, \alpha_d), \beta = (\beta_0, \beta_1, \ldots, \beta_d) \in \cI_{N,d}$ the indices for $\Sym(\fX, N)$.
The value $\alpha_i$ (or $\beta_i$) represents the number of times which $E_i$ (or $A_i$) appears in the tensor product.
In particular, we denote by $e_i = (0, \ldots, 0, 1, 0, \ldots, 0)$ the vector of length $d+1$ with $1$ appearing at the $i$-th position (starting from $0$).
Naturally $N e_i$ is understood as $(0,\ldots, 0, N, 0, \ldots, 0) \in \cI_{N,d}$.
The adjacency matrices and primitive idempotents of $\Sym(\fX,N)$ are thus denoted by $A_{\beta}$ and $E_{\alpha}$ respectively, where $\alpha, \beta \in \cI_{N,d}$.
The eigenmatrices and cosine matrices of $\Sym(\fX, N)$ can be expressed by multivariate Krawtchouk polynomials, more specifically the Aomoto-Gelfand series.
We will give the formula in the next subsection.

\subsection{Multivariate Krawtchouk polynomials}

In this subsection we discuss the relation between extension of commutative association scheme and multivariate Krawtchouk polynomials of Griffiths type (in complex coefficients).
Here we mostly follow the framework in \cite{MR2925407}.
In the end of this section we will discuss the relation among the multivariate Krawtchouk polynomials exhibited here and those in \cite{genest_multivariate_2013,MR2925407,mizukawa_n_2004}.

\begin{definition}
	Let $\nu$ be a nonzero real number.
	Let $P, \wt{P}$ be two $(d+1) \times (d+1)$ real matrices.
	And let $U$ be a $(d+1) \times (d+1)$ matrix with complex entries.
	We denote by $\cT_d$ be the set of $4$-tuples $(\nu, P, \wt{P}, U)$ satisfying the following assumptions.
	\begin{enumerate}
		\item $P = \diag(p_0, \ldots, p_d)$ and $\wt{P} = \diag(\wt{p}_0, \ldots, \wt{p}_d)$ are diagonal matrices and $p_0 = \wt{p}_0 = 1/\nu$.
		\item The entries in first row and the first column (with index $0$) of $U = (u_{i,j})_{0 \leq i, j \leq d}$ are $1$.
		\item
		      \begin{equation}\label{eqn:demurring}
			      \nu P U \wt{P} U^\dagger = I_{d+1}
		      \end{equation}
	\end{enumerate}
	For convenience, we use the shorthand $p = (p_0, p_1, \ldots, p_d)$ and $\wt{p} = (\wt{p}_0, \wt{p}_1, \ldots, \wt{p}_d)$.
\end{definition}

The $d$-variate Krawtchouk polynomials $\cK(n, \wt{n}) = \cK(n, \wt{n}, N, U)$ for $n, \wt{n} \in \cI_{N,d}$ are defined by the following expansion.
\begin{equation}
	\prod_{i=0}^d \paren{1 + \sum_{j=1}^d u_{i,j} z_j}^{\wt{n}_i} = \sum_{n \in \cI_{N,d}} \binom{N}{n} \cK(n, \wt{n}) z_1^{n_1} \cdots z_d^{n_d}.
\end{equation}

Mizukawa and Tanaka \cite{mizukawa_n_2004} gave an explicit formula by means of Aomoto-Gelfand hypergeometric series.
\begin{equation}
	\cK(n, \wt{n}) = \sum_{A = (a_{i,j}) \in \cM_{d,N}}
	\frac{\prod_{j=1}^d (-n_j)_{\sum_{i=1}^d a_{i,j}} \prod_{i=1}^d (-\wt{n}_i)_{\sum_{j=1}^d a_{i,j}}}{(-N)_{\sum_{i,j=1}^d a_{i,j}}} \prod_{i,j=1}^d \frac{\omega_{i,j}^{a_{i,j}}}{a_{i,j}!},
\end{equation}
where 
\begin{equation}
(x)_r = 
\begin{cases}
1, & r=0, \\
x(x+1) \cdots (x+r-1), & r>0,
\end{cases}
\end{equation}
is the standard Pochhammer symbol, $\omega_{i,j} = 1 - u_{i,j}$, and $\cM_{d,N}$ is the set of $d \times d$ non-negative integer matrices with entry sum $N$.

\begin{remark}
	In \cite{MR2925407}, the matrices $P$ and $\wt{P}$ are complex and the identity is $\nu P U \wt{P} U^\top = I_{d+1}$.
	Here $P$ and $\wt{P}$ are real and the identity is $\nu P U \wt{P} U^\dagger = I_{d+1}$.
	The reason is that the orthogonality is given by symmetric bilinear form there while here it is given by sesquilinear form.
	They are two types of different complex Krawtchouk polynomials with different parametrization.
	Note that the order of $n$ and $\wt{n}$ here in $\cK(n, \wt{n})$ is the reverse of the order of $-\alpha$ and $-\beta$ in the formula in \cite{mizukawa_n_2004}.
	Therefore for $\alpha, \beta \in \cI_{N,d}$, the entries in the cosine matrix of $\Sym(\fX, N)$ is given by $c_{\alpha, \beta} = \cK(\beta, \alpha, N, C) = \cK(\alpha, \beta, N, C^\top)$, where $C$ is the cosine matrix of $\fX$.
\end{remark}

The details of multivariate Krawtchouk polynomials will be deferred to the appendix.
The proofs are similar to those in \cite{MR2925407}.
We include them here in order to make the paper self-contained and exhibit the difference between two types of complex Krawtchouk polynomials.

\section{State transfer on extensions of association scheme}\label{sec:overdose}

Consider $M = \sum_{i=1}^d w_i A_{(N-1)e_0 + e_i}$ with the restriction that $w_{i'} = \overline{w_i}$, namely $M$ is a Hermitian.
Note that $e^{-\ii t M}$ is still contained in the Bose-Mesner algebra.
We can rewrite $M$ in the basis $E_{\alpha}, \cond{\alpha \in \cI}$ by
\begin{align}
\begin{split}
	M & = \sum_{i=1}^d w_i \sum_{\alpha \in \cI} P_{\alpha, (N-1)e_0 + e_i} E_{\alpha}          \\
	  & = \sum_{\alpha \in \cI} \paren{\sum_{i=1}^d w_i P_{\alpha, (N-1)e_0 + e_i}} E_{\alpha}.
\end{split}
\end{align}
The eigenvalues $\lambda_{\alpha}$ of $M$ for $E_{\alpha}$ can be further expressed by eigenvalues in $\fX$, namely
\begin{align}
\begin{split}
	\lambda_{\alpha} & = \sum_{i=1}^d w_i P_{\alpha, (N-1)e_0 + e_i}                                                                        \\
	                 & = \sum_{i=1}^d w_i \sum_{j=0}^d \alpha_j P_{j,i}                                                          \\
	                 & = \sum_{j=1}^d \alpha_j \paren{\sum_{i=1}^d w_i P_{j,i}} + \alpha_0 \paren{\sum_{i=1}^d w_i P_{0,i}}      \\
	                 & = N \paren{\sum_{i=1}^d w_i P_{0,i}} + \sum_{j=1}^d \alpha_j \paren{\sum_{i=1}^d w_i (P_{j,i} - P_{0,i})} \\
	                 & = N \paren{\sum_{i=1}^d w_i k_i} - \sum_{j=1}^d \alpha_j \paren{\sum_{i=1}^d w_i k_i (1 - c_{j,i})}.
\end{split}
\end{align}

Therefore $e^{-\ii tM}$ can be computed as follows.
\begin{align}
\begin{split}
	e^{-\ii t M} & = \sum_{\alpha \in \cI} e^{-\ii t \lambda_{\alpha}} E_{\alpha}                                                                                       \\
	             & = \sum_{\alpha \in \cI} e^{-\ii t \lambda_{\alpha}} \paren{\frac{1}{\abs{X}^N} \sum_{\beta \in \cI} Q_{\beta, \alpha} A_{\beta}}                     \\
	             & = \sum_{\beta \in \cI} \frac{1}{\abs{X}^N}\paren{\sum_{\alpha \in \cI} e^{-\ii t \lambda_{\alpha}} m_{\alpha} \overline{c_{\alpha,\beta}}} A_{\beta}
\end{split}
\end{align}

Since the $\ol{c_{\alpha, \beta}} = \ol{\cK(\alpha, \beta, N, C^\top)} = \cK(\alpha, \beta, N, C^\dagger)$, we apply \cref{eqn:nonutility} to the coefficient for $A_{\beta}$ in $e^{-\ii t M}$.
\begin{align}
\begin{split}
	f_{\beta}(t) & = \frac{1}{\abs{X}^N} \sum_{\alpha \in \cI} e^{-\ii t \lambda_{\alpha}} m_{\alpha} \cK(\alpha, \beta)                                                                                              \\
	             & = \frac{1}{\abs{X}^N} e^{-\ii t N \sum_{i=1}^d w_i k_i} \sum_{\alpha \in \cI} \binom{N}{\alpha} \cK(\alpha,\beta) \prod_{j=1}^d m_j^{\alpha_j} e^{\ii t \alpha_j \sum_{i=1}^d w_i k_i (1-c_{j,i})} \\
	             & = \frac{1}{\abs{X}^N} e^{-\ii t N \sum_{i=1}^d w_i P_{0,i}} \prod_{k=0}^d \paren{1 + \sum_{\ell=1}^d \overline{c_{\ell,k}} m_{\ell} e^{\ii t \sum_{i=1}^d w_i k_i (1-c_{\ell,i})} }^{\beta_k}      \\
	             & = \frac{1}{\abs{X}^N} e^{-\ii t N \sum_{i=1}^d w_i k_i} \prod_{k=0}^d \paren{1 + \sum_{\ell=1}^d \overline{c_{\ell,k}} m_{\ell} e^{\ii t \sum_{i=1}^d w_i k_i (1-c_{\ell,i})} }^{\beta_k}.
\end{split}
\end{align}
For ease of notation, we let
\begin{equation}
	z_{\ell} = z_{\ell}(t) = e^{\ii t \sum_{i=1}^d w_i k_i (1 - c_{\ell,i})},
\end{equation}
and let $p_k = 1 + \sum_{\ell=1}^d \ol{c_{\ell,k}} m_{\ell} z_{\ell}$.

As long as $p_k = 0$, the coefficient $f_{\beta}(t)$ vanishes on $\beta \in \cI$ with $\beta_k \neq 0$.
In order to achieve perfect state transfer or fractional revival, we want $f_\beta(t)$ to vanish for as many $\beta$ as possible.

We would like to remark that once the values of $z_{\ell}(t)$ ($\abs{z_{\ell}} = 1$), $\ell = 1,2, \ldots, d$ are given, one can always find $w_i$, $i = 1,2, \ldots, d$ such that $z_\ell = e^{-\ii t \sum_{i=1}^d w_i (P_{\ell,i} - P_{0,i})}$.
Consider the linear equation $\widetilde{P} w = b$, where $\widetilde{P}_{\ell,i} = P_{0,i} - P_{\ell,i} $ for $1 \leq \ell, i \leq d$, and $w = (w_1,w_2, \ldots, w_d)^\top$, $b = \frac{1}{\ii t} (\arg z_1, \arg z_2, \ldots, \arg z_d)^\top$.
Since $\widetilde{P}$ can be obtained by operating an elementary row transformation on $P$, which is of full rank, the matrix $\widetilde{P}$ is also of full rank.
Therefore the linear equation $\widetilde{P} w = b$ always has solutions.

Perfect state transfer on $\Sym(\fX, N)$ can be understood in two regards.
On one hand, fix a base vertex $x_0 \in X^N$ with the indicator vector $| x_0 \rangle = \delta_{x_0}$.
Then the evolution gives $e^{-\ii tM} |x_0 \rangle = \sum_{x \in X^N} c_x(t) |x \rangle$ for some functions $c_x(t)$.
On the other hand, set $| Y_{\beta} \rangle = \frac{1}{\sqrt{k_{\beta}}} A_\beta | x_0 \rangle$ for $\beta \in \cI$.
Then $e^{-\ii tM} | Y_{N e_0} \rangle = \sum_{\beta \in \cI} f_\beta(t) \sqrt{k_\beta}  | Y_\beta \rangle$.
So we have quantum walks among sites $Y_\beta$ with $\beta \in \cI$.

The computation in this section can be summarized as the following theorem.

\begin{theorem}
	Let $\fX = (X, \set{R_i}_{i=0}^d)$ be a commutative association scheme of class $d$.
	Let $A_{\beta} \cond{\beta \in \cI}$ be the adjacency matrices in $\Sym(\fX, N)$ in the N-th symmetric tensor product of $\fX$.
	Consider the quantum walks on a Hermitian matrix $M = \sum_{i=1}^d w_i A_{(N-1)e_0 + e_i}$.
	Then $e^{-\ii t M} = \sum_{\beta \in \cI} f_{\beta}(t) A_{\beta}$, where
	\begin{equation*}
		f_{\beta}(t) = \frac{1}{\abs{X}^N} e^{-\ii t N \sum_{i=1}^d w_i k_i} \prod_{k=0}^d \paren{1 + \sum_{\ell=1}^d \overline{u_{\ell,k}} m_{\ell} e^{\ii t \sum_{i=1}^d w_i k_i (1-c_{\ell,i})} }^{\beta_k}.
	\end{equation*}
	In particular, suppose the quantum walk starts at a site $\ket{x_0}$, then it will not occur at a site $\ket{Y_{\beta}} = \frac{1}{\sqrt{k_{\beta}}} A_\beta \ket{x_0}$ with $\beta_k \neq 0$ at time $t$ as long as
	\begin{equation*}
		1 + \sum_{\ell=1}^d \overline{c_{\ell,k}} m_{\ell} e^{\ii t \sum_{i=1}^d w_i k_i (1-c_{\ell,i})}  = 0.
	\end{equation*}
\end{theorem}

Next we compute the matrix representation $B_M$ of the matrix $M$ projected into the invariant subspace spanned by $\ket{Y_\beta}, \beta \in \cI$.

Firstly we have 
\begin{align}
	A_{(N-1)e_0 + e_i} A_{\beta} = & \sum_{j=0}^d \beta_j p_{ij}^j A_\beta + \sum_{s \neq t} (\beta_t + 1) p_{is}^t A_{\beta - e_s + e_t}. 
\end{align}

Note that 
\begin{align}
	\bra{x_0} A_\gamma^\dagger A_\alpha \ket{x_0} &= \sum_{\xi} \Tr p_{\gamma' \alpha}^\xi A_\xi \ket{x_0} \bra{x_0}
	= p_{\gamma' \alpha}^{N e_0} 
	= k_\gamma \delta_{\gamma}^\alpha. 
\end{align}

Therefore
\begin{align}
\begin{split}
	&\bra{Y_\gamma} M \ket{Y_\beta} \\
	=& \frac{1}{\sqrt{k_\gamma} k_\beta} \sum_{i=1}^d w_i \bra{x_0} A_\gamma^\dagger A_{(N-1)e_0 + e_i} A_\beta \ket{x_0} \\
	=& \frac{k_\gamma}{k_\beta} \sum_{i=1}^d w_i \paren{\sum_{j=0}^d \beta_j p_{ij}^j \delta_\gamma^\beta + \sum_{s \neq t} (\beta_t+1) p_{is}^t \delta_\gamma^{\beta - e_s + e_t}} \\
	=& 
	\begin{cases}
		\sum_{i=1}^d w_i \sum_{j=0}^d \beta_j p_{ij}^j, & \gamma = \beta, \\
		\sqrt{\beta_s (\beta_t + 1)} \sum_{i=1}^d w_i p_{is}^t, & \gamma = \beta - e_s + e_t, (s \neq t) \\
		0, & \otherwise,
	\end{cases}
\end{split}
\end{align}
where $M = \sum_{i=1}^d w_i A_{(N-1)e_0 + e_i}$, and $\delta_\gamma^\beta$ is the Kronecker delta function. 

Hence 
\begin{align}
	(B_M)_{\beta, \gamma} = 
	\begin{cases}
		\sum_{i=1}^d w_i \sum_{j=0}^d \beta_j p_{ij}^j, & \gamma = \beta, \\
		\sqrt{\beta_s (\beta_t + 1)} \sum_{i=1}^d w_i p_{is}^t, & \gamma = \beta - e_s + e_t, (s \neq t) \\
		0, & \otherwise.
	\end{cases}
\end{align}

\section{Quantum walks on symmetric tensor product of directed $n$-gon and multiple state transfer over arbitrary distances} \label{sec:incompliant}

In this section we apply the extension of association schemes to  the association scheme of directed $n$-gon. 
Namely we shall consider the symmetric tensor product of directed $n$-gon. Then we consider the quantum walks of the graph associated with the association scheme and show the interesting phenomena can be observed we call multiple state transfer over arbitrary distances.

\subsection{The association scheme of directed \texorpdfstring{$n$}{n}-gon}\label{subsec:cocozelle}
As shown in \cref{empl:hoove}, the first eigenmatrix of the directed $n$-gon is given by
\begin{equation}
	P = \begin{bmatrix}
		1      & 1           & 1              & \cdots & 1               \\
		1      & \zeta       & \zeta^2        & \cdots & \zeta^{n-1}     \\
		\vdots & \vdots      & \vdots         & \vdots & \zeta^{2(n-1)}  \\
		1      & \zeta^{n-1} & \zeta^{2(n-1)} & \cdots & \zeta^{(n-1)^2}
	\end{bmatrix},
\end{equation}
where $\zeta$ is the $n$-th root of unity, and the second eigenmatrix and the cosine matrix are given by $Q = \overline{P}$, and $C = P$.
In other words, $c_{\ell, k} = \zeta^{\ell k}$ for $0 \leq \ell, k \leq n-1$.
The valencies $k_j$ and multiplicities $m_i$ are all equal to $1$.
Here the entries of the eigenmatrices can be understood as the evaluation of orthogonal polynomials $z^{k}, (k=0,1,\ldots, n-1)$ with the weight function taking point mass at $\zeta^0, \zeta^1, \ldots, \zeta^{n-1}$ on the unit circle.

Suppose $z_{\ell}(\tau_k) = e^{k \ell \frac{2 \pi \ii}{n}} = \zeta^{\ell k}$ for $1 \leq \ell \leq n-1$, then $p_0 = p_1 = \cdots = p_{k-1} = p_{k+1} = \cdots = p_{n-1} = 0$ since $PQ = n I$.
In this case, $f_{\beta}(\tau_k)$ concentrates on $N e_{k}$.
In other words, we have perfect state transfer from $\ket{Y_{N e_0}}$ to $\ket{Y_{N e_k}}$ at time $t = \tau_k$.

One solution is given by $w_j = \frac{1}{\zeta^{-j} - 1}$ and $\tau_k = \frac{2 k \pi}{n}$ for $1 \leq j,k \leq n-1$.
Namely we have multiple perfect state transfers among $\ket{Y_{N e_0}}$, $\ket{Y_{N e_1}}$, $\ket{Y_{N e_2}}$, $\ldots$, $\ket{Y_{N e_{n-1}}}$.
Similar to the case in \cref{subsec:philosophicojuristic}, we also have multiple perfect state transfer among $\ket{Y_{\beta}}$, $\ket{Y_{\sigma(\beta)}}$, $\ket{Y_{\sigma^2(\beta)}}$, $\ldots$, $\ket{Y_{\sigma^{n-1}(\beta)}}$, where $(\sigma(\beta))_i = \beta_{i-1 \pmod n} \cond{0 \leq i \leq n-1}$. 

Therefore we obtain the following theorem. 

\begin{theorem}
	For the association scheme of directed $n$-gon $\mathfrak{Z}_n$ ,
	let $A_{\beta} \cond{\beta \in \cI}$ be the adjacency matrices in $\Sym(\mathfrak{Z}_n, N)$ in the N-th symmetric tensor product of $\mathfrak{Z}_n$.
	Consider the quantum walks on a Hermitian matrix $M = \sum_{i=1}^d w_i A_{(N-1)e_0 + e_i}$ with $w_i = \frac{1}{\zeta^{-i}-1}$ where $\zeta$ is $n$-th root of unity. 

	Suppose the quantum walk starts at a site $\ket{Y_{N e_0}} = \ket{x_0}$, then perfect state transfer occurs at the site $\ket{Y_{{N e_k}}}$ at time $\tau_k = \frac{2k\pi}{n}$ for $k = 1,2, \ldots, n-1$. 
	In other words, we have multiple perfect state transfer among extreme points of a simplex. 
\end{theorem}

Recall that $B_M$ is the matrix representation of $M$ projected into the invariant subspace spanned by $\ket{Y_\beta}, \beta \in \mathcal{I}$. 
Note that for the case of directed $n$-gon, the matrix $B_M$ is given by
\begin{align}
	(B_M)_{\beta, \gamma} = 
	\begin{cases}
		\sqrt{\beta_s (\beta_t +1)} w_{t-s \mod n}, & \gamma = \beta - e_s + e_t, \\
		0, & \otherwise. 
	\end{cases}
\end{align}

We remark that the spectrum of $B_M$ is integral when taking $w_i = \frac{1}{\zeta^{-i}-1}, i = 1,2, \ldots, n-1$ where $\zeta$ is the $n$-th root of unity. 
Consider the matrix $R = \sum_{i=1}^{n-1} w_i A_i$, where $A_i = Z^i$ with $Z$ being the circulant matrix. 
The eigenvalues of $R$ are $\sum_{i=1}^{n-1} w_i \zeta^{ik}$ for $k = 0, 1, \ldots, n-1$, namely $0, -1, -2, \ldots, -(n-1)$. 
Note that $M = R \otimes I \otimes \cdots \otimes I + I \otimes R \otimes \cdots \otimes I + \cdots + I \otimes \cdots \otimes I \otimes R$. 
Therefore the eigenvalues of $M$ are $\sum_{i=1}^N \eta_i$ with $\eta_i \in \set{0,-1,-2,\ldots, -(n-1)}$. 
Consequently the eigenvalues of $B_M$ are $0 \gamma_0 - \gamma_1 - 2\gamma_2 - \cdots - (n-1)\gamma_{n-1}$ with $\gamma_0 + \gamma_1 + \cdots + \gamma_{n-1} = N$.

\subsection{Explicit examples: $n=2$}
\begin{figure}[htbp]
	\centering
	\begin{subfigure}[b]{0.2\textwidth}
		\centering
		\includegraphics[width=\textwidth]{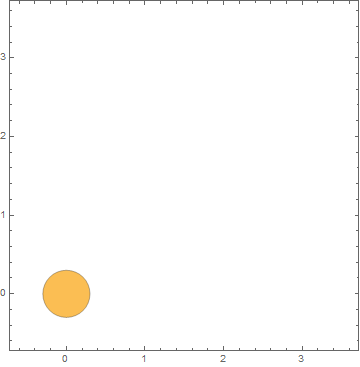}
		\caption{$t=0$}
		\label{fig:gon2dt0}
	\end{subfigure}
	\begin{subfigure}[b]{0.2\textwidth}
		\centering
		\includegraphics[width=\textwidth]{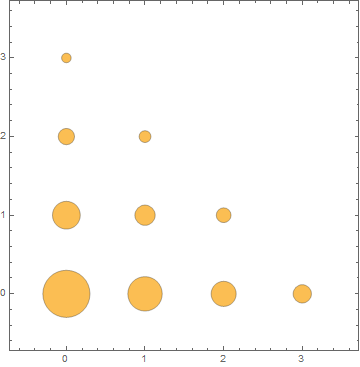}
		\caption{$t=2\pi/9$}
		\label{fig:gon2dt1}
	\end{subfigure}
	\begin{subfigure}[b]{0.2\textwidth}
		\centering
		\includegraphics[width=\textwidth]{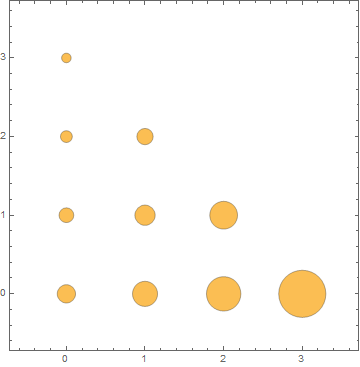}
		\caption{$t=4\pi/9$}
		\label{fig:gon2dt2}
	\end{subfigure}\\
	\begin{subfigure}[b]{0.2\textwidth}
		\centering
		\includegraphics[width=\textwidth]{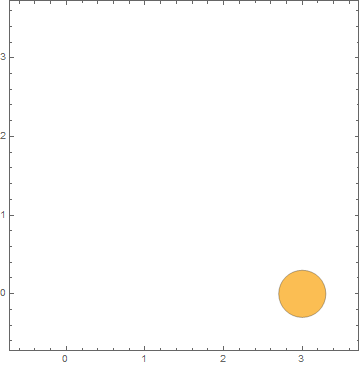}
		\caption{$t=2\pi/3$}
		\label{fig:gon2dt3}
	\end{subfigure}
	\begin{subfigure}[b]{0.2\textwidth}
		\centering
		\includegraphics[width=\textwidth]{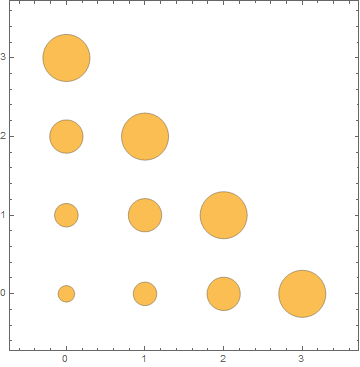}
		\caption{$t=\pi/2$}
		\label{fig:gon2dt4}
	\end{subfigure}
	\begin{subfigure}[b]{0.2\textwidth}
		\centering
		\includegraphics[width=\textwidth]{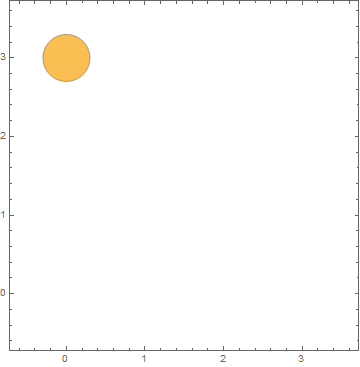}
		\caption{$t=4\pi/3$}
		\label{fig:gon2dt5}
	\end{subfigure}
	\caption{Multiple state transfer among extreme points of a $2$-simplex}
	\label{fig:gon2d}
\end{figure}

We shall here the concrete expression for $n=2$ in order to clarify the relationship to the bivariate Krawtchouk polynomials and show the figures of the corresponding quantum walks.
For the $N$-th symmetric tensor product of the directed 3-gon Sym($\mathfrak{Z}_3,N$), its adjacency matrices are given by 
\begin{equation}
C_{m,n}:=A_{N-m-n,m,n}=\frac{1}{\binom{N}{N-m-n ,m,n}}\sum_{\sigma \in S_N}A_{i_{\sigma(1)}}\otimes A_{i_{\sigma (2)}}\otimes \cdots \otimes  A_{i_{\sigma (N)}}
\end{equation}
with $N-m-n,m,n$ be the number of occurrences of $A_0,A_1,A_2$ respectively. Indeed, the set of the adjacency matrices 
\begin{equation}
\{ C_{m,n}~|~0\le m+n\le N\} =\{ A_{\ell ,m,n}~|~(\ell ,m,n)\in \cI_{N, 2}\}
\end{equation}
forms the commutative association scheme and its Bose-Mesner algebra is e.g. given by 
\begin{align}
\begin{split}
C_{1,0}C_{m,n}&=(N+1-m-n)C_{m,n-1}\\
&+(m+1)C_{m+1,j}+(n+1)C_{m-1,n+1},\\
C_{0,1}C_{m,n}&=(N+1-m-n)C_{m-1,n}\\
&+(n+1)C_{m,n+1}+(m+1)C_{m+1,n-1}.
\end{split}
\end{align}
From the above relation, one finds that
\begin{align}\label{eqn:bivariate-scheme}
C_{m,n}=p_{m,n}(C_{1,0},C_{0,1})
\end{align}
with 
\begin{align*}
p_{m,n}(\lambda_{x,y}^1,\lambda _{x,y}^2)=\binom{N}{N-m-n,m,n}G_{m,n}(x,y),
\end{align*}
where 
\begin{align}
\begin{split}
&\lambda_{x,y}^1= N+\sqrt{3}\ii ( \zeta y-\zeta ^{-1} x),\\
&\lambda_{x,y}^2= N+\sqrt{3}\ii (\zeta  x-\zeta ^{-1}y),\\
&G_{m,n}(x,y)= \sum_{0\le i+j+k+\ell \le N} \frac{(-m)_{i+j}(-n)_{k+\ell }(-x)_{i+k}(-y)_{j+\ell }}{(-N)_{i+j+k+\ell }}\frac{u_1^iv_1^ju_2^kv_2^{\ell }}{i!j!k!\ell !}
\end{split}
\end{align}
and $u_1=v_2=1-\zeta ,u_2=v_1=1-\zeta ^{-1 }$ and $\zeta $ is a cubic root of unity.
The expression \eqref{eqn:bivariate-scheme} implies that this association scheme is a bivariate $P$-polynomial association scheme \cite{bannai2023multivariate,bernard2022bivariate}.
As is mentioned in the previous section, the polynomials $\{ G_{m,n}(x,y)\}$ are bivariate Krawtchouk polynomials of the Griffith type whose orthogonality is given as follows:
\begin{equation}
\sum_{0\le x+y \le N} \omega _{x,y}G_{m_1,n_1}(x,y) \overline{G_{m_2,n_2}(x,y)}= \frac{\delta_{m_1,n_1}\delta_{m_2,n_2}}{3^N\binom{N}{N-m_1-n_1,m_1,n_1}}
\end{equation}
with $\omega _{x,y}=\binom{N}{N-x-y,x,y}p^xq^y(1-p-q)$ and $p=q=\frac{1}{3}$. It should be remarked here that the bivariate Krawtchouk polynomials given here are orthogonal polynomials of complex coefficients although their variables are still real (integer). \\
Since $A_1=Z,A_2=Z^2=A_1^{\top}$ and $A_0=E$, $C_{1,0}=C_{0,1}^{\top }$ and 
\begin{equation}
M= w _1 C_{1,0} +w _2 C_{0,1} \quad \left( w_1=\frac{1}{\zeta^{-1} -1 },w_2=\frac{1}{\zeta -1 }\right)
\end{equation}
is a Hermitian matrix of size $3^N$. In the manner given in the previous subsection, $M$ is projected into the matrix $B_M$ of size $\binom{N+2}{2}$ which is related with the recurrence relation of the orthonormal bivariate Krawtchouk polynomials:
\begin{align}
\begin{split}
&(w_1\lambda_{x,y}^1+w_2\lambda_{x,y}^2)\tilde{G}_{m,n}(x,y)\\
&=w_1\sqrt{(N-m-n)(m+1)}\tilde{G}_{m+1,n}(x,y)\\
&+w_1\sqrt{m(n+1)}\tilde{G}_{m-1,n+1}(x,y)\\
&+w_2\sqrt{(N-m-n)(n+1)}\tilde{G}_{m,n+1}(x,y)\\
&+w_2\sqrt{n(m+1)}\tilde{G}_{m+1,n-1}(x,y)\\
&+w_1\sqrt{(N+1-m-n)n}\tilde{G}_{m,n-1}(x,y)\\
&+w_2\sqrt{(N+1-m-n)m}\tilde{G}_{m-1,n}(x,y)
\end{split}
\end{align}
with $\tilde{G}_{m,n}(x,y)=\sqrt{3^N\binom{N}{N-m-n,m,n}}G_{m,n}(x,y)$.
For instance, when $N=3$, $B_M$ is given as follows:  
\begin{equation}\label{eqn:traducian}
{\tiny	\begin{bmatrix}
		0            & \sqrt{3} w_1 & \sqrt{3} w_2 & 0           & 0            & 0           & 0           & 0           & 0           & 0           \\
		\sqrt{3} w_2 & 0            & w_1          & 2 w_1       & \sqrt{2} w_2 & 0           & 0           & 0           & 0           & 0           \\
		\sqrt{3}w_1  & w_2          & 0            & 0           & \sqrt{2}w_1  & 2w_2        & 0           & 0           & 0           & 0           \\
		0            & 2w_2         & 0            & 0           & \sqrt{2}w_1  & 0           & \sqrt{3}w_1 & w_2         & 0           & 0           \\
		0            & \sqrt{2}w_1  & \sqrt{2}w_2  & \sqrt{2}w_2 & 0            & \sqrt{2}w_1 & 0           & \sqrt{2}w_1 & \sqrt{2}w_2 & 0           \\
		0            & 0            & 2w_1         & 0           & \sqrt{2}w_2  & 0           & 0           & 0           & w_1         & \sqrt{3}w_2 \\
		0            & 0            & 0            & \sqrt{3}w_2 & 0            & 0           & 0           & \sqrt{3}w_1 & 0           & 0           \\
		0            & 0            & 0            & w_1         & \sqrt{2}w_2  & 0           & \sqrt{3}w_2 & 0           & 2w_1        & 0           \\
		0            & 0            & 0            & 0           & \sqrt{2}w_1  & w_2         & 0           & 2w_2        & 0           & \sqrt{3}w_1 \\
		0            & 0            & 0            & 0           & 0            & \sqrt{3}w_1 & 0           & 0           & \sqrt{3}w_2 & 0
	\end{bmatrix}}.
\end{equation}
In the quantum walks on the oriented graph with the adjacency matrix $B_M$ (equivalently $M$), one can observe perfect state transfer from $(0,0)$ to $(N,0)$ at $t=\frac{2\pi}{3}$ and another perfect state transfer from $(0,0)$ to $(0,N)$ at $t=\frac{4\pi}{3}$ which is depicted in \cref{fig:gon2d}. This is nothing but the multiple perfect state transfer.

The extension of association scheme can also be applied to other association schemes to obtain various fractional revivals.
We will put them in the appendices.

\section{Conclusion}

In this paper we investigate quantum walks on extensions of commutative association schemes.
The transition amplitude for these walks are shown to be explicitly calculated by multivariate Krawtchouk polynomials of Griffith type. In particular, in case of directed $n$-gon, the complex Krawtchouk polynomials play important roles and we found
multiple perfect state transfer among extreme points in a simplex of arbitrary dimension $d$ and arbitrary distance $N$ that cannot be seen in the quantum walks on unoriented graphs.
Examples of fractional revival on subsimplexes are given as well.

Based on these results, we propose two problems for further research.

\begin{problem}
What is the smallest (regarding the number of vertices and edges) graph which carries multiple perfect state transfer among $m$ vertices of pairwise distance at least $N$?
\end{problem}
Note that \cref{sec:incompliant} provides examples of graphs with $\Theta(N^{m-1})$ vertices and of bounded degree $\Theta(m^2)$, which is a relatively sparse graph.
In \cite{coutinho_quantum_2019}, the problem of the smallest unweighted graph with perfect state transfer between two vertices of distance $N$ was investigated.
The problem is easy if one considers perfect state transfer between two vertices on a weighted graph.
However it is not trivial when $m > 2$.

\section*{Acknowledgement}
The research of ST is supported by JSPS KAKENHI (Grant
Numbers 19H01792).
DZ thanks Shanghai Jiao Tong University for providing scholarships.
The authors would like to thank P.-A. Bernard, N. Cramp\'{e}, P. Iliev, L. Vinet, and M. Zaimi for discussions and sharing their results.

\bibliographystyle{plain}

\bibliography{symmetric_tensor}

\appendix

\section{Multivariate Krawtchouk polynomial}

In this section we establish the multivariate Krawtchuok polynomials in complex variables.
The framework of the proofs follows \cite{MR2925407}.
The main difference between these two types of polynomials is that the orthogonality relation is given by symmetric bilinear form in \cite{MR2925407}, while here it is given by the sesquilinear form.

We recall the setting of the multivariate Krawtchouk polynomial.

\begin{definition}
	Let $\nu$ be a nonzero real number.
	Let $P, \wt{P}$ be two $(d+1) \times (d+1)$ real matrices.
	And let $U$ be a $(d+1) \times (d+1)$ matrix with complex entries.
	We denote by $\cT_d$ be the set of $4$-tuples $(\nu, P, \wt{P}, U)$ satisfying the following assumptions.
	\begin{enumerate}
		\item $P = \diag(p_0, \ldots, p_d)$ and $\wt{P} = \diag(\wt{p}_0, \ldots, \wt{p}_d)$ are diagonal matrices and $p_0 = \wt{p}_0 = 1/\nu$.
		\item The entries in first row and the first column (with index $0$) of $U = (u_{i,j})_{0 \leq i, j \leq d}$ are $1$.
		\item
		      \begin{equation}
			      \nu P U \wt{P} U^\dagger = I_{d+1}
		      \end{equation}
	\end{enumerate}
\end{definition}

\subsection{Cartan subalgebras of \texorpdfstring{$\Lalgsl_{d+1}(\CC)$}{sl {d+1} (C)}}

For $i,j \in \set{0,1,\ldots, d}$, we denote by $\epsilon_{i,j}$ the matrix with with $1$ at the $(i,j)$-th entry and $0$ at all other entries.
Then the standard Cartan subalgebra $H$ of $\Lalgsl_{d+1}(\CC)$ is spanned by the basis $\set{\phi_1, \phi_2, \ldots, \phi_d}$, where $\phi_i = \epsilon_{i,i} - \frac{1}{d+1} I_{d+1}$ for $i=0,1,2, \ldots, d$.
Note that $\phi_0 + \phi_1 + \cdots + \phi_d = 0$.
We take a $(d+1) \times (d+1)$ matrix
\begin{equation}
	R = \wt{\theta} \wt{P} U^\dagger,
\end{equation}
where $\wt{\theta} \in \RR$ such that $\abs{\det(R)} = 1$.
Since $\nu P U \wt{P} U^\dagger = I_{d+1}$, we have $R^{-1} = \theta P U$, where $\theta = \frac{\nu}{\wt{\theta}}$.
Consider the automorphism $\Ad_R$ on $\Lalgsl_{d+1}(\CC)$ given by $\Ad_R(\beta) = R \beta R^{-1}$ for every $\beta \in \Lalgsl_{d+1}(\CC)$.
The images of $\phi_k$ and $e_{i,j}$ under this automorphism $\Ad_R$ are denoted by $\wt{\phi}_k$ and $\wt{\epsilon}_{i,j}$ respectively, namely $\wt{\phi}_k = R \phi_k R^{-1}$ and $\wt{\epsilon}_{i,j} = R \epsilon_{i,j} R^{-1}$ for $i,j, k \in \set{0,1,\ldots, d}$.
Then
\begin{equation}
	\wt{H} = \Span\set{\wt{\phi}_1, \wt{\phi}_2, \ldots, \wt{\phi}_d}
\end{equation}
is a conjugated Cartan subalgebra of $\Lalgsl_{d+1}(\CC)$.
We can expand $\wt{\phi}_i$ for $i \in \set{1,2, \ldots, d}$ in the basis $\set{\phi_j, \epsilon_{k, \ell}}$, $j,k,\ell \in \set{0,1,\ldots, d}, j \neq 0, k \neq \ell$ of $\Lalgsl_{d+1}(\CC)$ as follows:
\begin{equation}
	\wt{\phi}_i = \nu \sum_{k \neq \ell} p_i \wt{p}_k \ol{u_{i,k}} u_{i,\ell} e_{k,\ell} + \sum_{j=1}^d p_i (\nu \wt{p}_j \abs{u_{i,j}}^2 - 1) \phi_j.
\end{equation}
Similarly we can express $\phi_i$ for $i \in \set{1,2, \ldots, d}$ in the another basis $\set{\wt{\phi}_j, \wt{\epsilon}_{k, \ell}}$, $j,k,\ell \in \set{0,1,\ldots, d}, j \neq 0, k \neq \ell$ of $\Lalgsl_{d+1}(\CC)$:
\begin{equation}
	\phi_i = \nu \sum_{k \neq \ell} \wt{p}_i p_k u_{k,i} \ol{u_{\ell,i}} \wt{\epsilon}_{k,\ell} + \sum_{j=1}^d \wt{p}_i (\nu p_j \abs{u_{j,i}}^2 - 1) \wt{\phi}_j.
\end{equation}

There is an antiautomorphism $\mathfrak{a}$ on $\Lalgsl_{d+1}(\CC)$ which preserves both $H$ and $\wt{H}$ given as follows:
\begin{equation}
	\mathfrak{a}(\beta) = \wt{P}\beta^\top \wt{P}^{-1}, \beta \in \Lalgsl_{d+1}(\CC).
\end{equation}
It follows from definition that $\mathfrak{a}$ is an involution.
The following lemma explains why $\mathfrak{a}$ preserves both $H$ and $\wt{H}$.
\begin{lemma}
	\begin{align}
		\mathfrak{a}(\phi_i)         & = \phi_i,                                   & \mathfrak{a}(\wt{\phi}_i)         & = \wt{\phi}_i,                         & i \in \set{0,1,2, \ldots, d},        \\
		\mathfrak{a}(\epsilon_{i,j}) & = \frac{\wt{p}_j}{\wt{p}_i} \epsilon_{j,i}, & \mathfrak{a}(\wt{\epsilon}_{i,j}) & = \frac{p_j}{p_i} \wt{\epsilon}_{j,i}, & i \neq j \in \set{0,1,2, \ldots, d}.
	\end{align}
\end{lemma}

By direct computation, one may verify that for $0 \leq i \neq j \leq d$,
\begin{equation}
	\epsilon_{i,j} = \frac{\brac{\phi_j, \brac{\phi_i, \brac{\phi_j, \wt{\phi}_0}}} - \brac{\phi_i, \brac{\phi_j, \wt{\phi}_0}} }{2\wt{p}_i}.
\end{equation}
Therefore we obtain the following lemma.
\begin{lemma}
	The two Cartan subalgebra together generate $\Lalgsl_{d+1}(\CC)$.
\end{lemma}

\subsection{The action of \texorpdfstring{$\Lalgsl_{d+1}(\CC)$}{sl d+1 (C)} on the polynomials}

Let $z_0, z_1, \ldots, z_d$ be mutually commutative complex variables.
We use the shorthand $z = (z_0, z_1, \ldots, z_d)$.
The $\CC$-algebra $\CC[z]$ is the algebra of all complex polynomials in $z_0, z_1, \ldots, z_d$.

Consider the following representation $\rho: \Lalgsl_{d+1}(\CC) \to \Lalggl (\CC[z])$ given by
\begin{align}
	\rho(\epsilon_{i,j}) & = z_i \partial_{z_j},                      \\
	\rho(\phi_i)         & = z_i \partial_{z_i} - z_j \partial_{z_j}.
\end{align}
Then $\CC[z]$ becomes a $\Lalgsl_{d+1}(\CC)$-module.
Let $V_N \leq \CC[z]$ be the space of homogeneous polynomials of degree $N$, namely $V_N = \Span\set{z^n \mid n \in \cI_{N,d}}$.
It is easy to verify that $\Lalgsl_{d+1}(\CC)$ preserves the total degree of the polynomials, hence $V_N$ is a $\Lalgsl_{d+1}(\CC)$-submodule.
In particular, we have
\begin{align}
	\phi_i \cdot z^n         & = \paren{n_i - \frac{N}{d+1}} z^n, & 0 \leq i \leq d,        \\
	\epsilon_{i,j} \cdot z^n & = \lambda_j z^{n + e_i - e_j},     & 0 \leq i \neq j \leq d.
\end{align}

Consider the change of variables $\wt{z} = (\wt{z}_0, \wt{z}_1, \ldots, \wt{z}_d)$ given by
\begin{align}
	\wt{z} = z R.
\end{align}
Then we have
\begin{align}
	\wt{\phi}_i \cdot \wt{z}^n    & = \paren{n_i - \frac{N}{d+1}} \wt{z}^n, & 0 \leq i \leq d,        \\
	\wt{\epsilon}_{i,j} \cdot z^n & = \lambda_j \wt{z}^{n + e_i - e_j},     & 0 \leq i \neq j \leq d.
\end{align}

Again we will express $\wt{z}$ by $z$.
\begin{align}
	\wt{z}_0 & = \wt{\theta} \sum_{j=0}^d \wt{p}_j z_j, \label{eqn:distilland}                                                                       \\
	\wt{z}_k & = \wt{\theta} \wt{p}_0 z_0 + \wt{\theta} \sum_{j=1}^d \ol{u_{k,j}} \wt{p}_j z_j, & k \in \set{1,2, \ldots, d}. \label{eqn:neurohumor}
\end{align}
Set $\omega_{i,j} = 1 - u_{i,j}$, and we get
\begin{align}\label{eqn:packmaker}
	\wt{z}_k & = \wt{z}_0 - \wt{\theta} \sum_{j=1}^d \wt{p}_j \ol{\omega_{k,j}} z_j, & k \in \set{1,2, \ldots, d}.
\end{align}

\subsection{A sesquilinear form}

We define a sesquilinear form $\lrangle{-,-}$ on $V_N$ by
\begin{align}
	\lrangle{z^n, z^m} & = \delta_{n,m} \frac{n!}{\wt{p}^n} \theta^N, & n,m \in \cI_{N,d}.
\end{align}
Here we use the convention that $\lrangle{\lambda g, h} = \lambda \lrangle{g,h} = \lrangle{g, \ol{\lambda} h}$ for $\lambda \in \CC$, $g,h \in V_N$.

Note that $\lrangle{\beta \cdot g, h} = \lrangle{g, \mathfrak{a}(\beta) \cdot h}$ for every $\beta \in \Lalgsl_{d+1}(\CC)$, $g, h \in V_N$.
Moreover, $\wt{z}^n, n \in \cI_{N,d}$ are also orthogonal with respect to this sesquilinear form.

\begin{lemma}
	For $n, m \in \cI_{N,d}$, we have
	\begin{align}
		\lrangle{\wt{z}^n, \wt{z}^m} & = \delta_{n,m} \frac{n!}{p^n} \wt{\theta}^N.
	\end{align}
\end{lemma}

\begin{proof}
	If $n \neq m$, then at least one coordinate of $n$ and $m$ are different, say $n_i \neq m_i$.
	Note that $(n_i - \frac{N}{d+1}) \lrangle{\wt{z}^n, \wt{z}^m} = \lrangle{\wt{\phi}_i \cdot \wt{z}^n, \wt{z}^m} = \lrangle{\wt{z}^n, \mathfrak{a}(\wt{\phi}_i) \cdot \wt{z}^m} = (m_i - \frac{N}{d+1}) \lrangle{\wt{z}^n, \wt{z}^m}$.
	Therefore $\lrangle{\wt{z}^n, \wt{z}^m} = 0$.
	Now suppose $n = m$.
	We prove the identity by induction on $N - n_0 = n_1 + n_2 + \cdots + n_d$.
	For $n_1 + n_2 + \cdots + n_d = 0$, namely $n = (N, 0, \ldots, 0)$,
	\begin{align}
		\lrangle{\wt{z}^n, \wt{z}^n}
		 & = \lrangle{\wt{z}_0^N, \wt{z}_0^N}                                                                                                     \\
		 & = \lrangle{\paren{\wt{\theta} \sum_{j=0}^d \wt{p}_j z_j}^N, \paren{\wt{\theta} \sum_{j=0}^d \wt{p}_j z_j}^N }                          \\
		 & = \wt{\theta}^{2N} \lrangle{ \sum_{m \in \cI_{N,d}} \frac{N!}{m!} \wt{p}^{m} z^m, \sum_{m \in \cI_{N,d}} \frac{N!}{m!} \wt{p}^{m} z^m} \\
		 & = \wt{\theta}^{2N} \sum_{m \in \cI_{N,d}} \frac{(N!)^2}{(m!)^2} \wt{p}^{2m} \frac{m!}{\wt{p}^m} \theta^N                               \\
		 & =  \wt{\theta}^N (\wt{\theta} \theta)^N N! \sum_{m \in \cI_{N,d}} \frac{N!}{m!} \wt{p}^m                                               \\
		 & = \frac{N!}{\wt{p}_0^N} \wt{\theta}^N.
	\end{align}
	Therefore the assertion holds for $n = m = (N, 0, \ldots, 0)$.
	Now suppose $N-n_0 = n_1 + n_2 + \cdots + n_d > 0$.
	In particular, $n_i > 0$ for some $i \in \set{1,2, \ldots, d}$.
	\begin{align}
		n_i \lrangle{\wt{z}^{n+e_0-e_i}, \wt{z}^{n+e_0-e_i}}
		 & = \lrangle{\wt{\epsilon}_{0,i} \cdot \wt{z}^n, \wt{z}^{n+e_0 - e_i}}             \\
		 & = \lrangle{\wt{z}^n, \mathfrak{a}(\wt{\epsilon}_{0,i}) \cdot \wt{z}^{n+e_0-e_i}} \\
		 & = \frac{p_i}{p_0} (n_0 + 1) \lrangle{\wt{z}^{n}, \wt{z}^{n}}.
	\end{align}
	Therefore
	\begin{align}
		\lrangle{\wt{z}^n, \wt{z}^n} = \frac{p_0}{p_j} \frac{n_i}{n_0+1} \lrangle{\wt{z}^{n+e_0-e_i}, \wt{z}^{n+e_0-e_i}}.
	\end{align}
	Hence the assertion holds by induction.
\end{proof}

\subsection{The Krawtchouk polynomials}

\begin{theorem}
	\begin{align}
		\frac{\wt{z}^{\wt{n}}}{\wt{\theta}^N}
		 & = N! \sum_{n \in \cI_{N,d}} \ol{\cK(n,\wt{n})} \frac{\wt{p}^n}{n!} z^n \label{eqn:phytosociology} \\
		\frac{z^n}{\theta^N}
		 & = N! \sum_{\wt{n} \in \cI_{N,d}} \cK(n,\wt{n}) \frac{p^{\wt{n}}}{\wt{n}!} {\wt{z}}^{\wt{n}}
	\end{align}
\end{theorem}

\begin{proof}
	We expand the left hand side of \cref{eqn:phytosociology} by \cref{eqn:packmaker} and the Multinomial Theorem.
	Here $a_{i,j}$ stands for the number of occurrences of $(-\wt{p}_j \ol{\omega_{k,j}} z_j)$ in the product of $\wt{n}_i$ terms.
	\begin{align}
		\frac{\wt{z}_i^{\wt{n}_i}}{\wt{\theta}^{\wt{n}_i}}
		 & = \sum_{(a_{i,j}) \in \cI_{\wt{n}_i,d}}
		\wt{n}_i ! \frac{\paren{\wt{z}_0 / \wt{\theta}}^{\wt{n}_i - \sum_{j=1}^d a_{i,j}}}{(\wt{n}_i-\sum_{j=1}^d a_{i,j})!}
		\prod_{j=1}^d \frac{(-\wt{p}_j \ol{\omega_{i,j}}z_j)^{a_{i,j}}}{a_{i,j}!}                                                    \\
		 & = \sum_{(a_{i,j})_{1 \leq j \leq d} \in \NN_0^{d}} \paren{\frac{\wt{z}_0}{\wt{\theta}}}^{\wt{n}_i - \sum_{j=1}^d a_{i,j}}
		(-\wt{n}_i)_{\sum_{j=1}^d a_{i,j}} \nonumber                                                                                 \\
		 & \phantom{=} \times
		\prod_{j=1}^d \frac{\ol{\omega_{i,j}}^{a_{i,j}}}{a_{i,j}!} (\wt{p}_i z_j)^{a_{i,j}} \label{eqn:Pantodon}
	\end{align}
	Note that in \cref{eqn:Pantodon} we use the fact that $(-\wt{n}_i)_{\sum_{j=1}^d a_{i,j}}$ vanishes once $\sum_{j=1}^d a_{i,j} > \wt{n}_i$.
	We multiply \cref{eqn:Pantodon} for $i \in \set{0,1,\ldots, d}$, and treat the terms with $\wt{z}_0$ separately.
	\begin{align}
		\frac{\wt{z}^{\wt{n}}}{\wt{\theta}^N}
		 & = \sum_{(a_{i,j}) \in \cM_{N,d}} \paren{\frac{\wt{z}_0}{\wt{\theta}}}^{N - \sum_{i,j=1}^d a_{i,j}}
		\prod_{i=1}^d (-\wt{n}_i)_{\sum_{j=1}^d a_{i,j}} \nonumber                                            \\
		 & \phantom{=} \times
		\prod_{j=1}^d (\wt{p}_j z_j)^{\sum_{i=1}^d a_{i,j}}
		\prod_{i,j=1}^d \frac{\ol{\omega_{i,j}}^{a_{i,j}}}{a_{i,j}!} \label{eqn:Excoecaria}
	\end{align}
	Then we expand the term with $\wt{z}_0^{N-s}$ again by \cref{eqn:distilland} and the Multinomial Theorem.
	Set $s = \sum_{i,j=1}^d a_{i,j}$.
	\begin{align}
		\paren{\frac{\wt{z}_0}{\wt{\theta}}}^{N - s}
		 & = \sum_{r \in \cI_{N-s,d}} (N-s)!
		\prod_{j=0}^d \frac{(\wt{p}_j z_j)^{r_j}}{r_j!}                                           \\
		 & = \sum_{r \in \cI_{N-s,d}} \frac{N!}{r_0! \prod_{j=1}^d (r_j + \sum_{i=1}^d a_{i,j})!} \\
		 & \phantom{=} \times
		\frac{\prod_{j=1}^d (-r_j - \sum_{i=1}^d a_{i,j})_{\sum_{i=1}^d a_{i,j}}}{(-N)_{\sum_{i,j=1}^d a_{i,j}}} \prod_{j=0}^d (\wt{p}_j z_j)^{r_j}.
	\end{align}
	We introduce new variables $n_0 = r_0$ and $n_j = r_j + \sum_{i=1}^d a_{i,j}$ for $j \in \set{1,2, \ldots, d}$.
	Then
	\begin{align}
		\paren{\frac{\wt{z}_0}{\wt{\theta}}}^{N - s}
		 & = \sum_{n \in \cI_{N,d}}
		\frac{N!}{n!} \frac{\prod_{j=1}^d (-n_j)_{\sum_{i=1}^d a_{i,j}}}{(-N)_{\sum_{i,j=1}^d a_{i,j}}} (\wt{p}_0 z_0)^{n_0} \prod_{j=1}^d (\wt{p}_j z_j)^{n_j - \sum_{i=1}^d a_{i,j}}. \label{eqn:naturalizer}
	\end{align}
	Finally we substitute \cref{eqn:naturalizer} into \cref{eqn:Excoecaria}.
	\begin{align}
		\frac{\wt{z}^{\wt{n}}}{\wt{\theta}^N}
		 & = \sum_{n \in \cI_{N,d}} \binom{N}{n} \prod_{j=0}^d (\wt{p}_j z_j)^{n_j} \nonumber \\
		 & \phantom{=} \times
		\sum_{(a_{i,j}) \in \cM_{N,d}} \frac{\prod_{i=1}^d (-\wt{n}_i)_{\sum_{j=1}^d a_{i,j}} \prod_{j=1}^d (-n_j)_{\sum_{i=1}^d a_{i,j}} }{(-N)_{\sum_{i,j=1}^d a_{i,j}}}
		\prod_{i,j=1}^d \frac{\ol{\omega_{i,j}}^{a_{i,j}}}{a_{i,j}!}                          \\
		 & = \sum_{n \in \cI_{N,d}} \binom{N}{n} \ol{\cK(n,\wt{n})} \wt{p}^n z^n              \\
		 & = N! \sum_{n \in \cI_{N,d}} \ol{\cK(n,\wt{n})} \frac{\wt{p}^n}{n!} z^n
	\end{align}
\end{proof}

Expand \cref{eqn:phytosociology} with \cref{eqn:neurohumor}, and apply the change of variables $z_j \mapsto \ol{z_j/\wt{p_j}}$ for $j \in \set{0,1,\ldots, d}$.
We get the generating function up to complex conjugation.

\begin{corollary}
	\begin{equation}\label{eqn:nonutility}
		\prod_{i=0}^d \paren{1 + \sum_{j=1}^d u_{i,j} z_j}^{\wt{n}_i} = \sum_{n \in \cI_{N,d}} \binom{N}{n} \cK(n, \wt{n}) z_1^{n_1} \cdots z_d^{n_d}.
	\end{equation}
\end{corollary}

Apply $\lrangle{z^n, -}$ to both sides of \cref{eqn:phytosociology}, we express the Krawtchouk polynomial in sesquilinear form.

\begin{corollary}
	\begin{align}
		\cK(n,\wt{n}) = \frac{1}{\nu^N N!} \lrangle{z^n, \wt{z}^{\wt{n}}}.
	\end{align}
\end{corollary}

Apply $\lrangle{-, \wt{z}^{\wt{k}}}$ to both sides of \cref{eqn:phytosociology}, we obtain the orthogonality relation.

\begin{corollary}
	For $\wt{n}, \wt{k} \in \cI_{N,d}$, we have
	\begin{align}
		N! \sum_{n \in \cI_{N,d}} \ol{\cK(n,\wt{n})} \cK(n,\wt{k}) \frac{\wt{p}^n}{n!} = \delta_{\wt{n},\wt{k}} \frac{\wt{n}!}{N! \nu^N p^{\wt{n}}},
	\end{align}
	and similarly for $n, k \in \cI_{N,d}$, we have
	\begin{align}
		N! \sum_{\wt{n} \in \cI_{N,d}} \ol{\cK(n,\wt{n})} \cK(k,\wt{n}) \frac{p^{\wt{n}}}{\wt{n}!} = \delta_{n,k} \frac{n!}{N! \nu^N \wt{p}^n}.
	\end{align}
\end{corollary}

\section{Relation to other fractional revivals}

In this section we apply the extension of association schemes to other association schemes.
They correspond to Perfect state transfer and Fractional revivals.

\subsection{The trivial association scheme of size \texorpdfstring{$2$}{2}}\label{subsec:philosophicojuristic}

In this subsection we revisit the classical construction of perfect state transfer over a long path with Krawtchouk weight.

The first eigenmatrix, the second eigenmatrix, and the cosine matrix of the trivial association scheme $\fX_2$ is given by
\begin{equation*}
	P = Q = C =
	\begin{bmatrix}
		1 & 1  \\
		1 & -1
	\end{bmatrix}.
\end{equation*}
It is in fact the Hadamard matrix $H_2$ of order $2$ as $C^\top C = 2 I$.
The $N$-th symmetric tensor product $\Sym(\fX_2, N)$ of $\fX_2$ is the binary Hamming association scheme of length $N$, in other words the association scheme of the $N$-hypercube.

If $0 = p_0 = 1 + z_1$, namely $z_1 = -1$, then $f_\beta(t)$ concentrates on $(0, N) \in \cI$.
In other words, we have perfect state transfer from $|Y_{(N, 0)} \rangle$ to $| Y_{(0, N)} \rangle$.
Moreover, note that
\begin{align}
	e^{-\ii \tau M} \ket{Y_{\beta}} & = e^{-\ii \tau M} \frac{1}{\sqrt{k_\beta}} A_\beta \ket{Y_{(N,0)}}   \\
	                                & = \frac{1}{\sqrt{k_\beta}} A_{\beta} e^{-\ii \tau M} \ket{Y_{(N,0)}} \\
	                                & = \frac{1}{\sqrt{k_\beta}}A_{\beta} \ket{Y_{(0,N)}}                  \\
	                                & = \ket{Y_{(\beta_1, \beta_0)}}.
\end{align}
So we have perfect state transfer from $\ket{Y_{(\beta_0, \beta_1)}}$ to $\ket{Y_{(\beta_1, \beta_0)}}$ as well for every $\beta = (\beta_0, \beta_1) \in \cI$.

\subsection{The ordered binary word association scheme}
In this part, we show that the ordered binary Hamming association scheme~\cite{MR1697147,miki_spin_2020} of depth $d$ can be obtained by taking the symmetric tensor product of small schemes, which we shall call ordered binary word association scheme $\OW(2,d)$ of depth $d$. 

Consider the following actions $g_j \cond{j = 2,3, \ldots, d}$ on $\fX_2^{\otimes d}$:
\begin{equation}
	g_j(A_{i_1} \otimes A_{i_2} \otimes \cdots \otimes A_{i_d}) = A_{i_1'} \otimes A_{i_2'} \otimes \cdots \otimes A_{i_d'}
\end{equation}
such that
\begin{equation}
	i_k' = \begin{cases}
		1 - i_k, & \textif i_j = 1, k = j-1, \\
		i_k,     & \otherwise.
	\end{cases}
\end{equation}
One can verify that $g_j \cond{j = 2,3, \ldots, d}$ are automorphism of the Bose-Mesner algebra of $\fX_2^{\otimes d}$.
The fusion scheme obtained by taking $G = \lrangle{g_2, g_3, \ldots, g_d}$ acting on $\fX_2^{\otimes d}$ is called the ordered binary word association scheme of depth $d$.

The first eigenmatrix of the ordered binary word association scheme $\OW(2,d)$ of depth $d$ is given by
\begin{equation}
	P = (P_{i,j})_{0\leq i,j \leq d} = \begin{bmatrix}
		1      & 1      & 2      & \cdots  & 2^{d-1}  \\
		1      & 1      & 2      & \cdots  & -2^{d-1} \\
		\vdots & \vdots & \vdots & \iddots & 0        \\
		1      & 1      & -2     & \iddots & \vdots   \\
		1      & -1     & 0      & \cdots  & 0
	\end{bmatrix},
\end{equation}
and the valencies $k_j$ and $m_i$ are given by $k_j = 2^{j-1}$ and $m_i = 2^{i-1}$ for $1 \leq i, j \leq d$ and $k_0 = m_0 = 1$.

The cosine matrix of $\OW(2,d)$ is given by
\begin{equation}
	C = (c_{i,j})_{0 \leq i,j \leq d} =
	\begin{bmatrix}
		1      & 1      & 1       & \cdots  & 1      \\
		1      & 1      & 1       & \iddots & -1     \\
		\vdots & \vdots & \iddots & \iddots & 0      \\
		1      & 1      & -1      & \iddots & \vdots \\
		1      & -1     & 0       & \cdots  & 0
	\end{bmatrix}
\end{equation}

We consider the $N$-th symmetric tensor power of $\OW(2,d)$, namely the ordered binary Hamming association scheme of length $N$ and depth $d$.

If we demand that $f_{\beta}(t)$ vanishes on $\beta \in \cI$ with $\beta_0 \neq 0$, then it requires that
\begin{align}
	0 = p_0 & = 1 + \sum_{\ell=1}^d \ol{c_{\ell,0}} m_{\ell} z_{\ell} \\
	        & = 1 + \sum_{\ell=1}^d 2^{\ell - 1} z_{\ell}.
\end{align}
Since $\abs{z_{\ell}} = 1$ for every $1 \leq \ell \leq d$, it forces that $z_{\ell} = 1$ for $\ell < d$ and $z_{d} = -1$ by triangle inequality.

If we demand that $f_{\beta}(t)$ vanishes on $\beta \in \cI$ with $\beta_k \neq 0 \cond{1 \leq k \leq d}$, then it requires that
\begin{align}
	0 = p_k & = 1 + \sum_{\ell=1}^d \ol{c_{\ell,k}} m_{\ell} z_{\ell}                \\
	        & = 1 + \sum_{\ell=1}^{d-k} 2^{\ell - 1} z_{\ell} - 2^{d-k+1} z_{d-k+1}.
\end{align}
Since $\abs{z_{\ell}} = 1$ for every $1 \leq \ell \leq d$, it forces that $z_{\ell} = 1$ for $\ell \leq d - k + 1$.

Therefore if $p_0 = 0$, then $p_2 = p_3 = \cdots = p_d = 0$.
In other words, $f_{\beta}(t)$ concentrates on $\beta = (N, 0, \ldots, 0)$.
Similarly if $p_k = 0$ for some $k \geq 1$, then $p_{k+1} = p_{k+2} = \cdots = p_d = 0$.
In other words, $f_{\beta}(t)$ concentrates on $\beta \in \cI$ with $\beta_k = \beta_{k+1} = \cdots = \beta_d = 0$ (a $k$-simplex).
Such a phenomenon was first observed in~\cite{miki_spin_2020}.

\begin{figure}[htbp]
	\centering
	\begin{subfigure}[b]{0.2\textwidth}
		\centering
		\includegraphics[width=\textwidth]{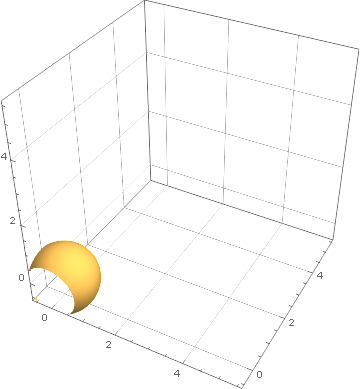}
		\caption{$t=0$}
		\label{fig:orderedHamming3dt0}
	\end{subfigure}
	\begin{subfigure}[b]{0.2\textwidth}
		\centering
		\includegraphics[width=\textwidth]{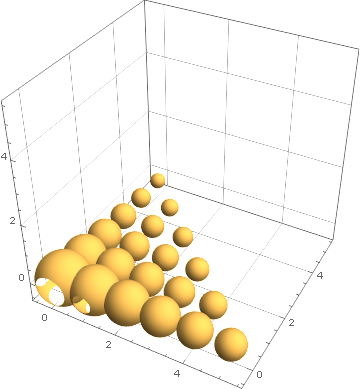}
		\caption{$t=\pi/16$}
		\label{fig:orderedHamming3dt1}
	\end{subfigure}
	\begin{subfigure}[b]{0.2\textwidth}
		\centering
		\includegraphics[width=\textwidth]{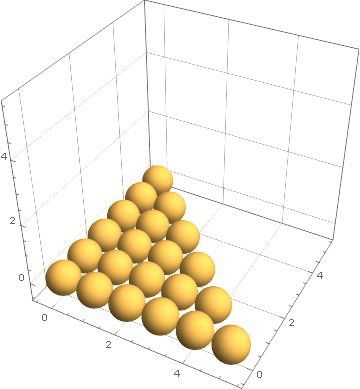}
		\caption{$t=\pi/8$}
		\label{fig:orderedHamming3dt2}
	\end{subfigure}
	\begin{subfigure}[b]{0.2\textwidth}
		\centering
		\includegraphics[width=\textwidth]{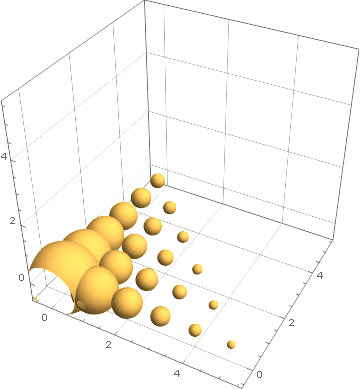}
		\caption{$t=3\pi/16$}
		\label{fig:orderedHamming3dt3}
	\end{subfigure}\\
	\begin{subfigure}[b]{0.2\textwidth}
		\centering
		\includegraphics[width=\textwidth]{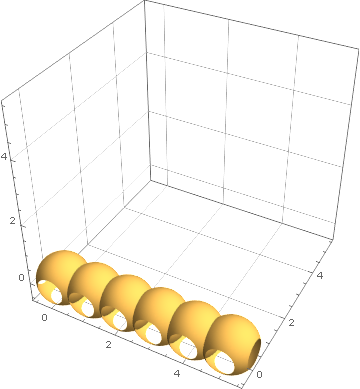}
		\caption{$t=\pi/4$}
		\label{fig:orderedHamming3dt4}
	\end{subfigure}
	\begin{subfigure}[b]{0.2\textwidth}
		\centering
		\includegraphics[width=\textwidth]{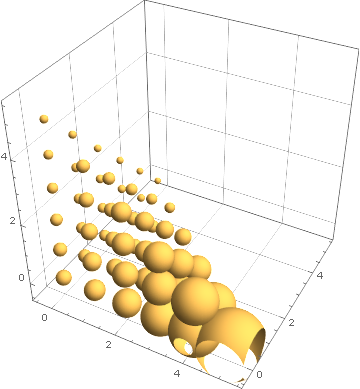}
		\caption{$t=\pi/3$}
		\label{fig:orderedHamming3dt5}
	\end{subfigure}
	\begin{subfigure}[b]{0.2\textwidth}
		\centering
		\includegraphics[width=\textwidth]{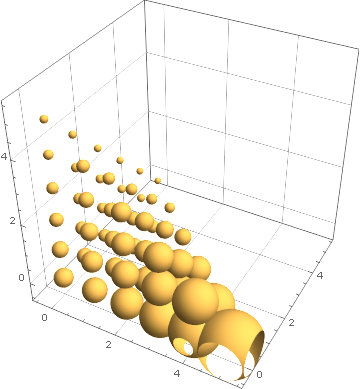}
		\caption{$t=5\pi/12$}
		\label{fig:orderedHamming3dt6}
	\end{subfigure}
	\begin{subfigure}[b]{0.2\textwidth}
		\centering
		\includegraphics[width=\textwidth]{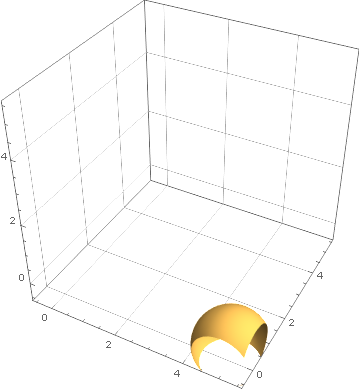}
		\caption{$t=\pi/2$}
		\label{fig:orderedHamming3dt7}
	\end{subfigure}
	\caption{Fractional revival on subsimplexes of a $3$-simplex}
	\label{fig:orderedHamming3d}
\end{figure}

\cref{fig:orderedHamming3d} illustrates the possibility of detecting the state at different sites and at various times through the size of the bubbles (based on the ordered Hamming association scheme of depth $3$ and length $5$).

\end{document}